\documentclass{article}%
\usepackage{spconf,graphicx}
\usepackage{amssymb,amsthm,amsfonts}
\usepackage[cmex10]{amsmath}
\usepackage{bm}
\usepackage{algorithm, algpseudocode, algorithmicx}

\title{Compressive Hyperspectral Imaging Using Progressive Total Variation}
\name{Simeon Kamdem Kuiteing$^\dagger$, Giulio Coluccia$^\star$, Alessandro Barducci$^\ddagger$, Mauro Barni$^\dagger$, Enrico Magli$^\star$}
\address{$^\dagger$ Dipartimento di Ingegneria dell'Informazione, Universit\`a di Siena, Italy\\
		 $^\star$ Dipartimento di Elettronica e Telecomunicazioni -- Politecnico di Torino, Italy\\
		 $^\ddagger$ Consiglio Nazionale delle Ricerche -- Istituto di Fisica Applicata ``Nello Carrara'', Firenze, Italy}

\newcommand{\mat}[1]{\ensuremath{\bm{\mathrm{#1}}}}

\newcommand{\F}{\ensuremath{\mat{F}}}
\newcommand{\y}{\ensuremath{\mat{y}}}
\newcommand{\e}{\ensuremath{\mat{e}}}
\newcommand{\E}{\ensuremath{\mat{E}}}

\newcommand{\Ph}{\ensuremath{\mat{\Phi}}}

\newcommand{\Y}{\ensuremath{\mat{Y}}}
\newcommand{\X}{\ensuremath{\mat{X}}}

\newcommand{\Nr}{\ensuremath{N_{\mathsf{R}}}}
\newcommand{\Nb}{\ensuremath{N_{\mathsf{B}}}}
\newcommand{\Nc}{\ensuremath{N_{\mathsf{C}}}}

\newcommand{\trasp}[1]{\ensuremath{#1 ^\mathsf{T}}}

\newcommand{\vect}[1]{\ensuremath{\mathrm{vec}\left \{ #1\right\}}}

\newcommand{\N}{\mathcal{N}}

\def\Ri{\mathbb{R}}

\begin{document}
\ninept
\maketitle
\begin{abstract}
Compressed Sensing (CS) is suitable for remote acquisition of hyperspectral images for earth observation, since it could exploit the strong spatial and spectral correlations, allowing to simplify the architecture of the onboard sensors. Solutions proposed so far tend to decouple spatial and spectral dimensions to reduce the complexity of the reconstruction, not taking into account that onboard sensors progressively acquire spectral rows rather than acquiring spectral channels. For this reason, we propose a novel progressive CS architecture based on separate sensing of spectral rows and joint reconstruction employing Total Variation. Experimental results run on raw AVIRIS and AIRS images confirm the validity of the proposed system.
\end{abstract}

\section{Introduction}
\label{sec:intro}

Compressed Senging (CS) \cite{candes2006cs, donoho2006cs} is a new signal acquisition paradigm, which takes advantage from the feature of many natural signals of being highly correlated. A high correlation entails the existence of a domain (usually defined by an integral transform) in which the signal is sparse, and only a small fraction of the transform coefficients are significantly different from zero. CS addresses the problem of collecting a number of measurements smaller than that required by Shannon theorem, but sufficient to allow the reconstruction of sparse signals with an arbitrarily low error.

A promising application of CS theory regards the remote acquisition of hyperspectral imagery for spaceborne and airborne earth observation. On one hand, hyperspectral images exhibit a strong correlation both in the spatial dimension and - more importantly - in the spectral dimension, so they fit perfectly the assumptions underlying CS theory. On the other hand, the use of CS techniques would allow to design sensors requiring a smaller memory buffer, fewer detectors, and a reduced volume of data to transmit.

The application of CS theory to hyperspectral image acquisition is not straightforward mainly due to the complexity of the reconstruction stage. For a fruitful application of CS, in fact, it is necessary that the redundancy of hyperspectral images in both the spatial and spectral dimension is exploited, in a truly 3D fashion. This implies that the measurement process and, more importantly, the reconstruction process are applied to 3D blocks of the image data cube of sufficient size, thus raising the problem of the computational complexity of the reconstruction step, which can quickly become unmanageable. See for example \cite{duarte2009kronecker}, where the properties of Kronecker products are exploited to cast a multidimensional problem to a single dimension vector whose size is the product of the sizes of each source dimension. Other approaches based on compressive projection principal component analysis \cite{fowler2009compressive} have been recently proposed in \cite{li2013classification} and \cite{ly2013reconstruction}, as well as a method relying on generalized tensor products \cite{li2013generalized}. See \cite{willettsparsity} for an overview of compressed sensing methods applied to hyperspectral imaging. 

A problem, common to virtually all proposed solutions to the computational complexity issue (noticeably \cite{coluccia2012progressive}), is that they work in a 2D + 1D fashion assuming that the two spatial dimensions, hereafter indicated as $x$-$y$, are acquired and processed together, and that the spectral dimension $\lambda$ is used in a second phase to progressively refine the reconstruction obtained from $x$-$y$ data, exploiting the correlation along the spectral dimension. Such an approach, however, does not take into account the way hyperspectral images are acquired by onboard sensors. In most cases, in fact, onboard systems are equipped with a linear array of sensors which, at a given time, acquires a spectral row ($x$ dimension at all wavelengths). The next spectral row is then acquired at the subsequent instant exploiting the motion of the satellite. This acquisition architecture is usually referred to as pushbroom configuration. As a result, the $y$ spatial dimension is essentially a time dimension, making it difficult to process first the images in the $x$-$y$ plane and add the spectral dimension in  a second time, since buffering the whole data cube is infeasible. A possible solution would be to apply 2D CS reconstruction to the $x$-$\lambda$ plane and use the $y$ dimension to refine the reconstruction. Unfortunately, previous attempts to do so failed to provide satisfactory results \cite{coluccia2012progressive}.

In this paper we propose a solution to this problem, {\em i.e.}, we consider the $x$-$\lambda$ plus $y$ configuration and investigate suitable reconstruction algorithms that are able to take advantage of correlations in all image dimensions. In particular, in light of the above issues, we propose a novel CS architecture, based on sensing over $x$-$\lambda$ {\em spectral rows} and reconstruction employing Total Variation (TV, \cite{candes2006ssr}) minimization, which is better suited to reconstruct images acquired by a satellite equipped with a sensor working in a pushbroom configuration. Unlike previous approaches, which reconstruct the data cube from a set of separately sensed spectral channels, the proposed algorithm requires separately sensed  spectral rows, which is compatible with the structure of pushbroom sensors. We show that the TV prior is effective at capturing the correlation within spectral rows, achieving a reconstruction quality very similar to that obtained by the simpler (but infeasible) conventional approach.

We validate the effectiveness of the proposed approach by testing it on data cubes acquired by the AVIRIS hyperspectral sensor and AIRS ultraspectral sounder. The results we obtain confirm the validity of our system that, thanks to the possibility of working on larger windows, achieves the same performance in both $x$-$y$ plus $\lambda$ and $x$-$\lambda$ plus $y$ configurations.


\section{Satellite onboard architectures}\label{sec:background}
\vspace*{-3mm}
With the term \emph{imaging spectrometers} we refer to instruments able to measure the energy emitted or reflected from an object as a function of two spatial and one spectral coordinate, originating 3D datasets called \emph{datacubes}. Unfortunately, only few 3D detectors exist, which have coarse spectral resolving power and poor efficiency, and are therefore unsuitable for the realization of spaceborne sensors. Modern imaging spectrometers employ 2D detector arrays, which collect a signal expressed in arbitrary digital units of energy as a function of three indices representing column, row, and exposure \cite{sellar2005comparison}. These raw data must be transformed into a standard coordinate/measurement system of at--sensor radiance, crosstrack position, along-track position, and wavelength (or wave number).
Imaging spectrometers used for earth observation can be categorized into classes based on two main criteria: the technique they adopt to perform spatial sampling and the sensors architecture utilized to induce spectral dispersion/discrimination. Techniques of spatial sampling that have been used in real remote sensing instruments are: whiskbroom, pushbroom, framing, and windowing. Evidently, due to the 3D nature of the signal to be collected the sampling scheme adopted in the spatial domain is not independent of the sampling scheme utilized for the remaining 1D spectral subset.
A whiskbroom scanning instrument employs a zerodimensional spatial field of view (FOV) that scans the object in both the along--track and the cross--track directions. Usually, this FOV covers the entire spectral interval to be sampled; \emph{i.e.}, a 1D detector array is adopted to observe all spectral channels with a single shot.
A pushbroom imaging spectrometer scans a 1D FOV in the along--track direction only, covering with a single acquisition the entire spectral range. A framing (also called staring) instrument employs a 2D FOV that remains fixed on the object during acquisition.
The term windowing will be used to describe the class of sensors that employ a 2D FOV that scans the target over the along—-track direction.
Three main techniques for separating spectral information produced by the observed source can be adopted in an aerospace sensor: wavelength filtering, spectral dispersion, and multiplexing. This last method can be applied into two distinct forms: interferential and dispersive multiplexing. Dispersive instruments use either a prism or a grating to obtain dispersion of the incoming light along a space direction, that is subsequently sampled by a matched detector. The most frequent type of multiplexing spectrometers encountered in aerospace sensors belongs to the class of Fourier--transform spectrometers (FTSs), \emph{i.e.}, two--beam interferometers such as the Michelson, Mach--Zehnder, and Sagnac optical configurations. Multiple–beam interferometers (aka \`{e}talons) such as the Fabry--Perot have signal collection abilities that are more similar to those of filtering instruments than of FTSs. The effects of the multiplexing architecture on the Signal--to--Noise (SNR) of the collected signal has been discussed in depth in \cite{barducci2010radiometric} and \cite{barducci2010theoretical}.
The most common type of imaging spectrometer used for the realization of remote sensing sensors is the pushbroom configuration associated with grating dispersion. Sensors belonging to this class adopt a 2D detector array that samples in a single shot a 2D domain composed by the across--track spatial coordinate on one direction, and the wavelength axis on the other one. A typical example of aerospace sensor having this architecture is the CHRIS-PROBA from the European Space Agency \cite{barducci2009investigating}. Therefore, we focus our investigation to assess and compressively sample 2D $x\lambda$ domains.
\subsection{CS Sensor Architecture}
\vspace*{-2mm}
\begin{figure}
\centering
\includegraphics[width=\columnwidth]{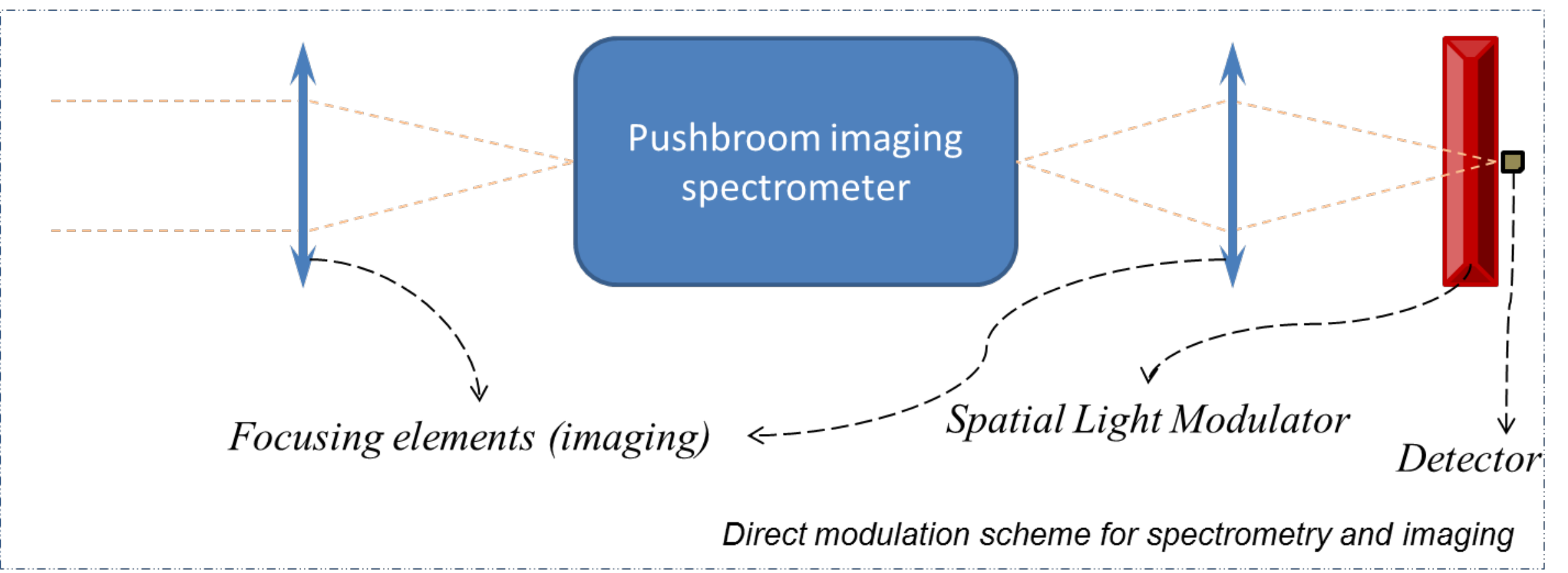}
\vspace*{-5mm}
\caption{\small Architecture of an ideal sensor utilizing the CS technology. The sensor modulates (spatial light modulator) the 2D domain output by the imaging spectrometer and focuses (integrates) the modulated domain on the single--point detector}
\label{fig:detector}
\end{figure}
The application of the CS technique to remote sensing requires a broadband light modulator that computes random projections of the observed image. It is important that these projections are implemented optically, thus avoiding the acquisition of the entire dataset to digitally perform the random linear combinations. Fig.~\ref{fig:detector} sketches the conceptual scheme of a CS hyperspectral imager operating in the pushbroom configuration.
The direct modulation scheme depicted in Fig.~\ref{fig:detector} adopts a single element detector, integrating the incoming radiation field as modulated by the Spatial Light Modulator (SLM). This last element is an electro--actuated 2D array of mirrors, crystals, or liquid crystals cells that modulates the available image before the acquisition performed by a single--element detector that integrates the image filtered by the SLM. It must be noticed that the availability of fast detectors and high frame-rate SLMs are critical points for any CS applications. Moreover, it is possible to build up a sensor with a SLM of lower frame rate, provided that a coarse resolution 2D array is utilized in the focal plane for parallelizing the CS of a mosaic of subimages.

\begin{figure}
\centering
\vspace*{-2mm}
\includegraphics[width=0.4\columnwidth]{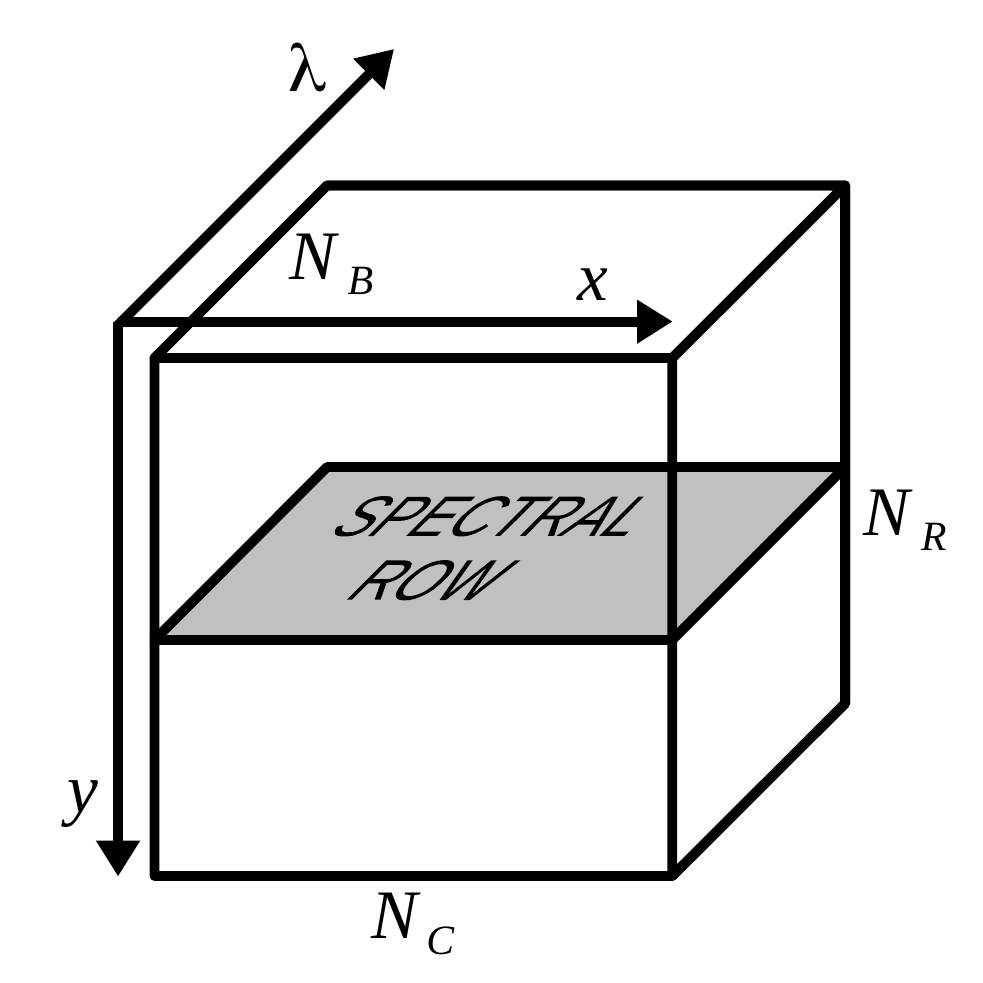}
\vspace*{-5mm}
\caption{Graphical representation of a $\Nr\times \Nc \times \Nb$ datacube. A spectral $x\lambda$ row is highlighted.}
\label{fig:hypercube}
\vspace*{-3mm}
\end{figure}

\vspace*{-3mm}
\section{Proposed technique}
\label{sec:proposed_technique}
\vspace*{-2mm}
Motivated by what stated in section~\ref{sec:background}, we propose an acquisition and reconstruction technique based on CS, capable of acquiring separate random projections of each spectral $x\lambda$ rows and reconstructing the entire data cube capturing the correlations in both spatial and spectral directions, with manageable complexity.

Referring to Fig.~\ref{fig:hypercube}, the hyperspectral image $\mathcal{F} \in \Ri^{N_{\mathsf{R}}\times N_{\mathsf{C}}\times N_{\mathsf{B}}}$ can be represented as a 3D collection of samples, where $x$ and $y$ represent spatial dimensions and $\lambda$ represents the spectral dimension. Hence, $\mathcal{F}$  can be considered as a collection of $N_{\mathsf{R}}$ spectral rows $\F^i = \mathcal{F}_{i,:,:}, i=1,\ldots,N_{\mathsf{R}}$, each consisting of a $N_{\mathsf{C}}\times N_{\mathsf{B}}$ matrix, \emph{i.e.}, $\mathcal{F}=[\F^1, \F^2, \ldots, \F^{\Nr}]$. We refer to this configuration as $x\lambda-y$ cube.

\subsection{Acquisition}
\vspace*{-2mm}

For what concerns the acquisition of the image, it consists in the collection of 
$\y_i\in\Ri^{M\times 1}$ of $M$ measurements for each spectral row as $ \y_i=\Ph^i\vect{\F^i}$,
where each sensing matrix $\Ph^i\in\Ri^{M\times \Nc\Nb}$ is taken as Gaussian i.i.d. and $M<\Nc\Nb$. For simplicity, $M$ is taken as the same value for all spectral rows. The measurements of all spectral rows are then collected in the matrix \Y. This setting is amenable to separate spatial reconstruction of each spectral
row. However, we expect that separate reconstruction does not yield a sufficiently accurate estimate of the original image, since it lacks modelling of vertical correlation. As a matter of fact, this inaccurate reconstruction will be the initialization point of our proposed reconstruction algorithm of section~\ref{sec:reconstruction}. The acquisition procedure is reported in Algorithm~\ref{alg:acquisition}.

\begin{algorithm}
\caption{Proposed acquisition algorithm}\label{alg:acquisition}
\begin{algorithmic}[1]
\Require the hyperspectral image $\mathcal{F}=[\F^1, \F^2, \ldots, \F^{\Nr}]$, the number of measurements per row $M$
\Ensure the measurement matrix \Y
\For{$i = 1$ to \Nr}
\State Draw $\Ph^i \in \Ri^{ M\times \Nc\Nb}$ s.t. $(\Ph^i)_{kj}\sim\N(0,1/M)$
\State$ (\Y)_i \gets \Ph^i\vect{\F^i}$
\EndFor
\State \Return \Y
\end{algorithmic}
\end{algorithm}

\vspace*{-2mm}
\subsection{Reconstruction with Iterative Total Variation (ITV)}
\label{sec:reconstruction}
\vspace*{-2mm}
The idea behind the iterative reconstruction is that if we can obtain a prediction of a spectral row $\F^i$, \emph{e.g.}, applying the operator  $\mathsf{P(\cdot,\cdot)}$ to rows $\F^{i-1}$ and $\F^{i+1}$ of some initial reconstruction, then we can cancel out the contribution of this predictor from the measurements of $\F^i$, and reconstruct only the prediction error instead of the full spectral row. If the prediction filter is accurate, the prediction error
is expected to be more compressible than the full signal, and the reconstruction will yield better results \cite{coluccia2012progressive}. In particular, the iterative procedure starts from the initial reconstruction $\mathcal{F}^{(0)}$ of all spectral rows. Even if this initial reconstruction can be obtained using several techniques, in our experiments we reconstruct each spectral row by solving, for each $i=1,\ldots,\Nr$, the following problem
\begin{align*}
&\mathcal{F}^{(0)}_{i,:,:} = \arg\min_{\X}\mathrm{TV}(\X)\quad\mathrm{s.t.}\quad\Ph\cdot\vect{\X} = (\Y)_i~,\ \text{where}\\
&\mathrm{TV}(\X) = \sum_{k,j}\sqrt{|(\X)_{k+1,j}-(\X)_{k,j}|^2+|(\X)_{k,j+1}-(\X)_{k,j}|^2}~.
\end{align*}
Since the TV is the sum of the magnitudes of the discretized gradient, seeking to minimize the TV norm relies on the assumption that the gradient of the spectral row is approximately sparse, hence the TV norm should be small. 
Then, for every row we first obtain its prediction from upper and lower rows estimated at previous iteration $\F_\mathsf{P} = \mathsf{P}\left(\mathcal{F}^{(n-1)}_{i-1,:,:},\mathcal{F}^{(n-1)}_{i+1,:,:}\right)$.
After that, we compute prediction error measurements as $\e_{\y} =
\y_i- \Ph^i\vect{\F_\mathsf{P}}$, using the prediction filter of \cite[Section III.C.2]{coluccia2012progressive}. We use $\e_{\y}$ to reconstruct the $i$-th
row summing the CS reconstruction of $\e_{\y}$ to $\F_\mathsf{P}$
\begin{align*}
\mathcal{F}^{(n)}_{i,:,:} &= \F_\mathsf{P} + \E_{\F}~,\ \text{where}\\
\E_{\F} &= \arg\min_{\E}\mathrm{TV}(\E)\quad\mathrm{s.t.}\quad\Ph\cdot\vect{\E} = \e_{\y}
\end{align*}
This process
is performed on all spectral rows, and is iterated until convergence.  The proposed iterative reconstruction scheme is shown in Algorithm \ref{alg:reconstruction}. 

\begin{algorithm}
\caption{ITV reconstruction algorithm}\label{alg:reconstruction}
\begin{algorithmic}[1]
\Require the measurement matrix \Y, the set of $\Ph^i$
\Ensure the estimation $\widehat{\mathcal{F}}$
\For{$i = 1$ to $\Nr$}
\State $\mathcal{F}^{(0)}_{i,:,:} \gets \arg\min_{\X}\mathrm{TV}(\X)\ \mathrm{s.t.}\ \Ph\cdot\vect{\X} = (\Y)_i$
\EndFor
\State $n \gets 0$
\Repeat
\State $n \gets n+1$
\For{$i = 1$ to $N_{\mathsf{BAND}}$}
\State  $\F_\mathsf{P} \gets \mathsf{P}\left(\mathcal{F}^{(n-1)}_{i-1,:,:},\mathcal{F}^{(n-1)}_{i+1,:,:}\right)$
\State $\y_\mathsf{P} \gets \Ph^i\cdot\vect{\F_\mathsf{P}}$
\State $\e_{\y} \gets \trasp{(\Y)_i} - \y_\mathsf{P}$
\State $\E_{\F} \gets \arg\min_{\E}\mathrm{TV}(\E)\ \mathrm{s.t.}\ \Ph\cdot\vect{\E} = \e_{\y}$
\State $\mathcal{F}^{(n)}_{i,:,:} \gets \F_\mathsf{P} + \E_{\F}$
\EndFor
\Until{Convergence is reached}
\State \Return $\mathcal{F}^{(n)}$
\end{algorithmic}
\end{algorithm}

\section{Results}
\label{sec:num_res}
\vspace*{-2mm}
\subsection{Dataset description}
\vspace*{-2mm}
Our simulations are mainly based on two hyperspectral images selected among those used as reference for onboard lossy compression in the ``multispectral and hyperspectral data compression'' working group of the Consultative Committee for Space Data Systems (CCSDS).
These images are commonly known as granule 9 ({\em gran9}) of AIRS and scene 0 (\emph{sc0}) of AVIRIS (Yellowstone).
AVIRIS is a spectrometer with 224 bands, and the size of this image is 512 lines and 680
pixels. AIRS is an ultraspectral sounder with 2378 spectral channels, used to create 3D maps of air and surface
temperature. In the CCSDS dataset, only 1501 bands are considered. The unstable channels have been removed as they have little or no scientific interest. The spatial size is 90 pixels and 135 lines. These scenes are widely used in literature so the comparisons with other
techniques are easier. Both are raw images i.e the output of the detector, without any
processing, calibration or denoising applied. Since our objective is to assess the potential of CS to manufacture
hyperspectral sensors, employing the raw image is a more realistic approach.

\vspace*{-2mm}
\subsection{Experimental results}
\vspace*{-2mm}

In order to evaluate the performance of our proposed algorithm, we present results from a set of experiments, given in terms of Mean Square Error (MSE) as a function of the percentage of measurements $M/N$, where $N = \Nr\Nc$ for the standard $xy-\lambda$ configuration and $N=\Nc\Nb$ for the $x\lambda-y$ configuration.

\begin{figure}
\begin{center}
 \includegraphics[width=0.8\columnwidth]{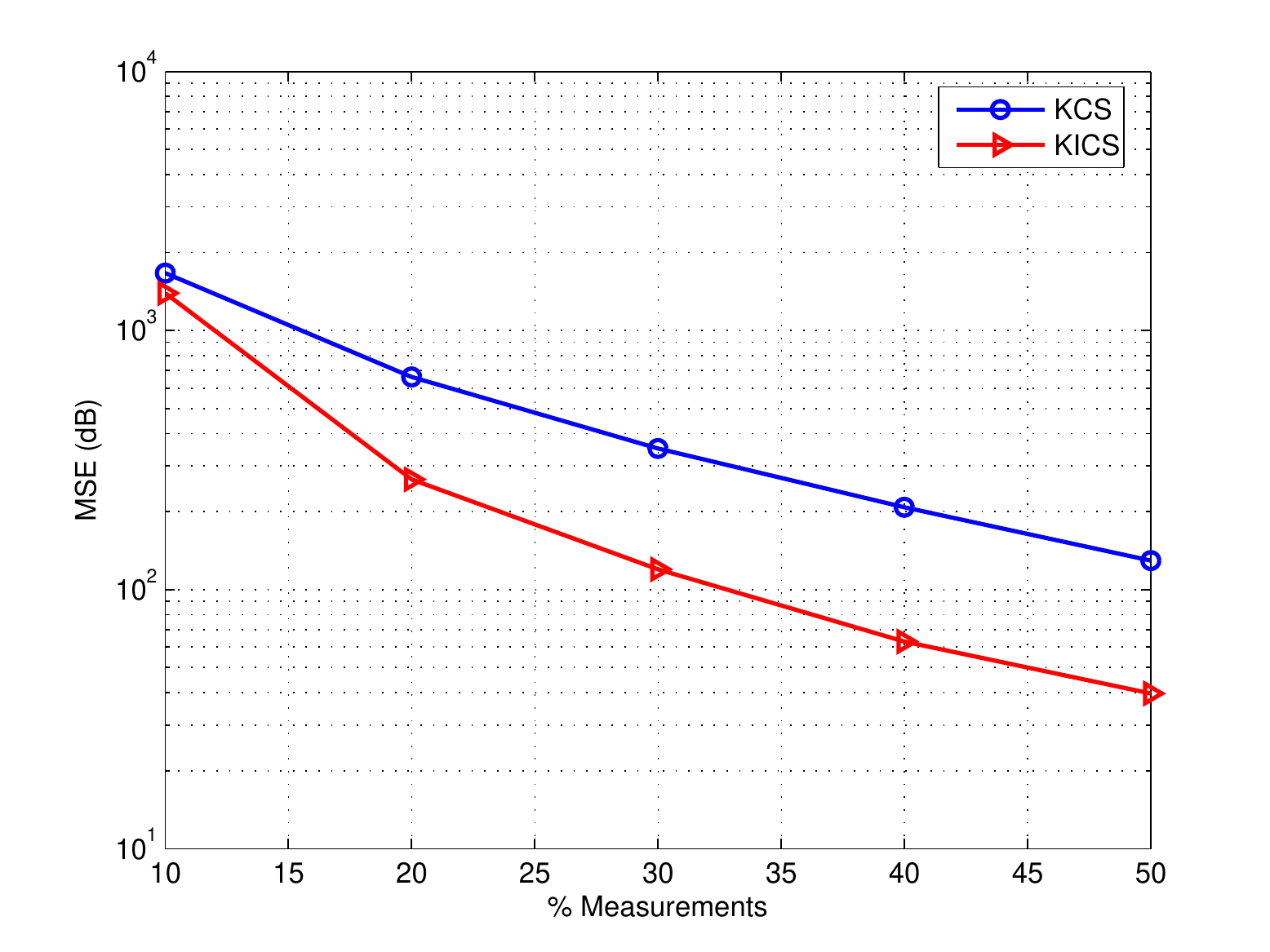}
\vspace{-5mm}
\caption{Reconstruction of AIRS scene: $32 \times 32$ $xy$ window.
}
\label{fig:ICASSP2014_FIG1}
\end{center}
\end{figure}

For the sake of comparison with our scheme, we report Fig.~\ref{fig:ICASSP2014_FIG1} which is obtained by considering the standard use of the hyperspectral data cube with $xy$ as spatial dimension and $\lambda$ as spectral one. 
To keep the computational complexity manageable, a $32\times 32$ spatial crop of the image across
all frequency bands was used. Fig.~\ref{fig:ICASSP2014_FIG1} shows the reconstruction of the AIRS scene performed by Kronecker Compressed Sensing (KCS) \cite{duarte2009kronecker} and those obtained through Kronecker-iterative compressed sampling (KICS) which relies on the iterative procedure described in \cite{coluccia2012progressive}, with the initial point computed by KCS. Note that both KCS and KICS perform signal recovery using the $\ell_1$-norm minimization process. As we can see, KICS provides quite good mean-squared error (MSE) values, but in the 3D reconstruction process, because of the large amount of data to deal with, KCS faces the computational problems related to $\ell_1$-norm minimization which provides the final image reconstruction.
As the complexity of $\ell_1$-norm minimization is cubic in the number of samples, increasing the dimension of the domain yields a very high complexity at the ground station.

\begin{figure}
\begin{center}
 \includegraphics[width=0.8\columnwidth]{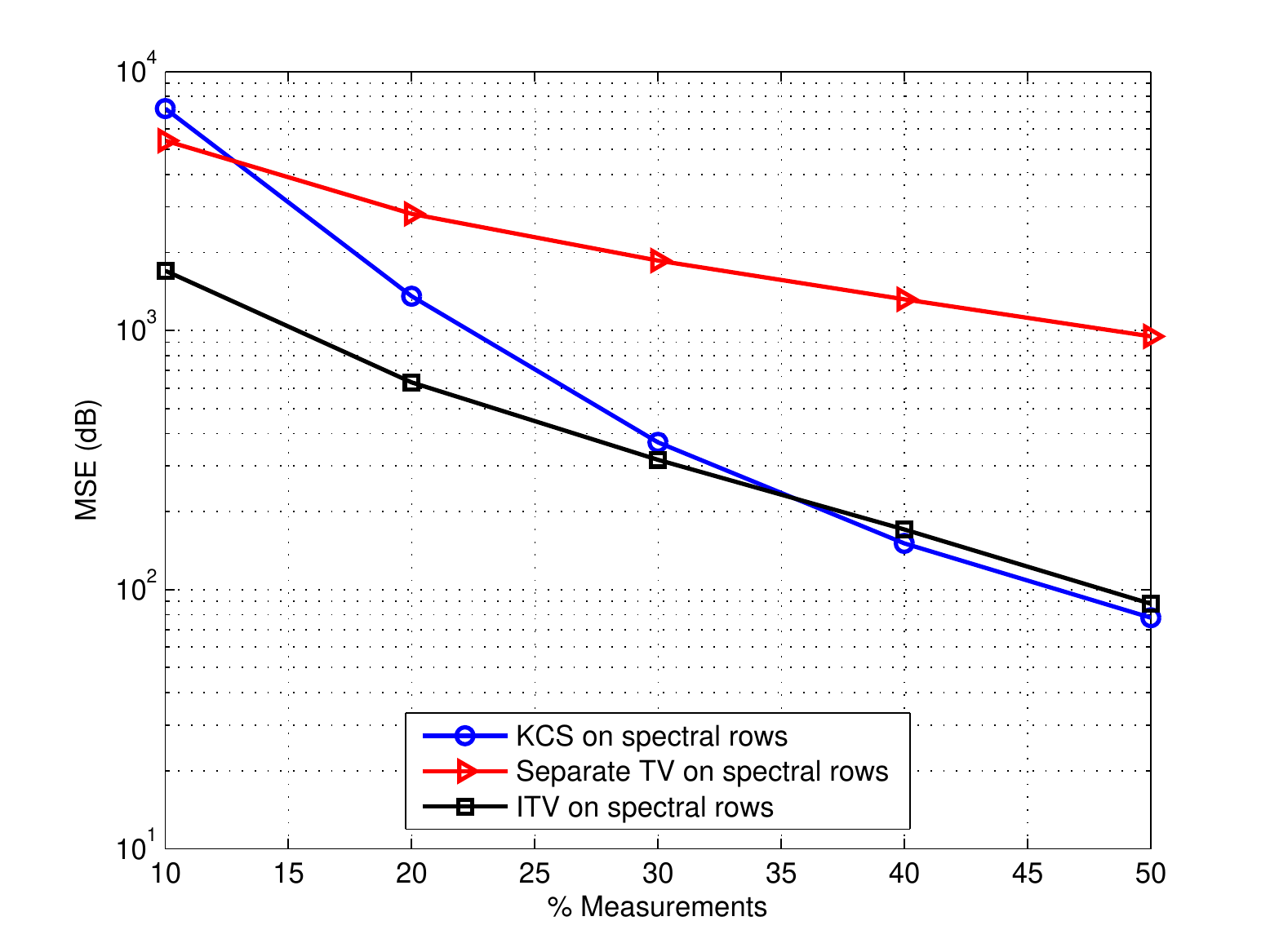}
 \vspace{-5mm}
\caption{Reconstruction of AIRS scene: $32 \times 32$ $x\lambda$ window.
}
\label{fig:ICASSP2014_FIG2}
\end{center}
\end{figure}

\begin{figure}
\begin{center}
 \includegraphics[width=0.8\columnwidth]{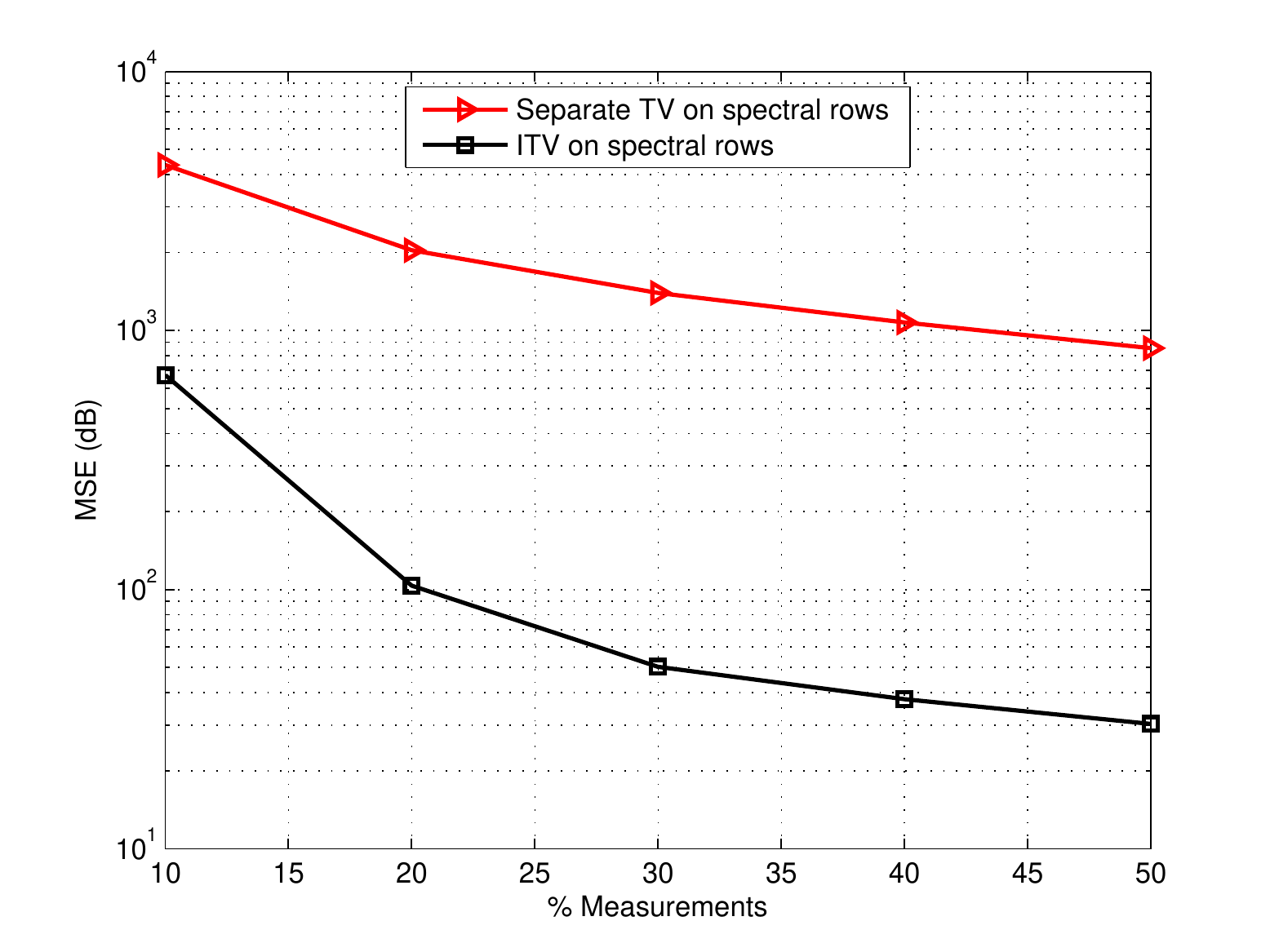}
 \vspace{-5mm}
\caption{Reconstruction of AIRS scene: $128 \times 128$ $x\lambda$ window.
}
\label{fig:ICASSP2014_FIG3}
\end{center}
\vspace*{-5mm}
\end{figure}

\begin{figure}
\begin{center}
 \includegraphics[width=0.8\columnwidth]{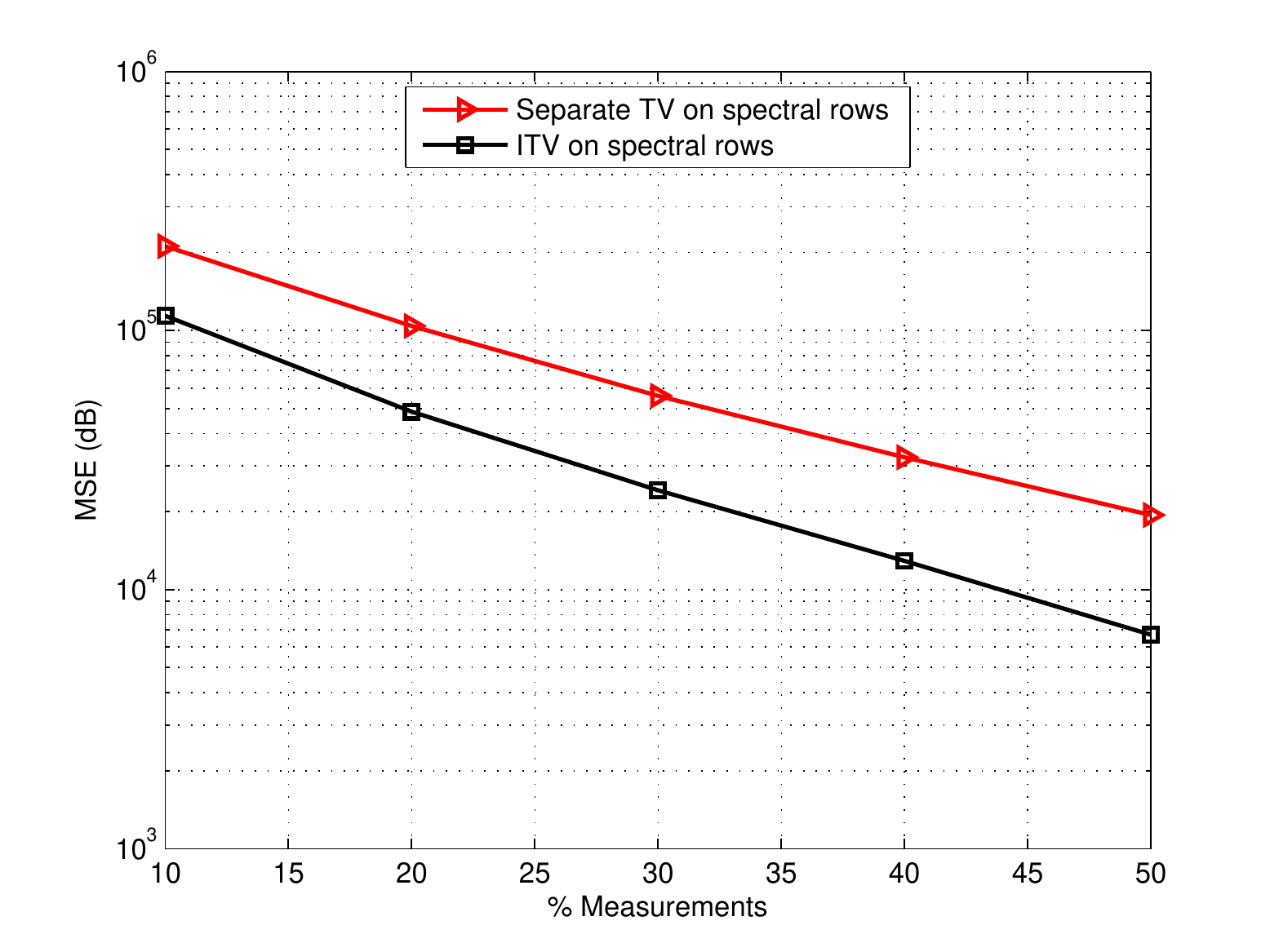}
 \vspace{-5mm}
\caption{Reconstruction of AVIRIS scene: $128 \times 128$ $x\lambda$ window.
}
\label{fig:ICASSP2014_aviris}
\end{center}
\vspace*{-5mm}
\end{figure}

In Figs.~\ref{fig:ICASSP2014_FIG2},~\ref{fig:ICASSP2014_FIG3} and ~\ref{fig:ICASSP2014_aviris}, we show the experiments with the $x\lambda-y$ cube. The cube is acquired using the procedure of Alg.~\ref{alg:acquisition} and reconstructed using ITV (Alg.~\ref{alg:reconstruction}). To have a fair comparison between KICS and our ITV algorithm, we focused on a small portion of the  hypercube, a $32 \times 32$ spectral $x\lambda$ rows and the whole vertical length ($y$). We repeated the experiment for 7 different windows along with their vertical dimension and averaged the MSE values obtained. Results are illustrated in Fig~\ref{fig:ICASSP2014_FIG2}. As can be seen, the 2D $x\lambda$ TV reconstruction yields very large MSE values, which is inappropriate for practical applications. It is worth mentioning that, in this scenario, KICS performed on the $x\lambda-y$ cube does not converge. The proposed ITV reconstruction algorithm converges in about 23 iterations and allows to improve significantly the MSE values.
ITV outperforms the KCS for $M<35\%$ and provides similar behavior to the KCS scheme for higher values of $M$. Moreover, the ITV algorithm allows to reduce drastically the computational complexity up to a factor of 12 with respect to the KCS schemes as shown in Tab.~\ref{tab:airs} (complexity results refer to a Matlab--based implementation running on a Windows operating system environment, equipped with Intel\textsuperscript{\textregistered} Core\texttrademark 2 Duo CPU T6500 @ 2.1 GHz processor and 4 GB Ram).
\begin{table}
\vspace*{-5mm}
\caption{{Computational time (min.): AIRS image}}
\centering
\begin{tabular}{|c|c|c|c|c|c|}
\hline
& \multicolumn{3}{|c|}{$32\times32$ crop} & \multicolumn{2}{|c|}{$128\times 128$ crop} \\
\hline

$M$ \% & \textit{KCS} & \textit{TV} & \textit{ITV} & \textit{TV} & \textit{ITV}\\
\hline
10  & 50 & 4 & 7 & 25 & 230 \\
\hline
30 &  98 & 8 & 14 & 43 & 400\\
\hline
50  & 245 & 12 & 21 & 63 & 650\\
\hline

\end{tabular}
\label{tab:airs}
\end{table} 

As a consequence, ITV could allow to reconstruct larger spatial-rows crops along with all their vertical dimension, a task very difficult to achieve with the KCS. Fig.~\ref{fig:ICASSP2014_FIG3} presents averaged results on three different $128\times 128$ $x\lambda$ windows along with all their vertical dimension. These results show that the bigger the $x\lambda$ window size the better the performance of ITV. On one hand, by using larger windows, ITV allows to improve significantly the MSE values with respect to the TV reconstruction while keeping the computational time at a very low level. For instance, for $M=30\%$, the TV takes about 40 minutes to reconstruct the entire 3D signal and a single iteration of the ITV reconstruction algorithm around 20 minutes. On other hand, by using a $128\times 128$ $x\lambda$ window along with all its $y$ vertical dimension, the KCS problem becomes computationally intractable and as a result the comparison with the ITV is impossible.
Experiments on the AVIRIS
image are shown in Fig.~\ref{fig:ICASSP2014_aviris} where the results lead to similar
observations to those made on the AIRS image.

\vspace*{-2mm}
\section{Conclusions}
\label{sec:conclusions}
\vspace*{-2mm}
In this paper, we proposed an architecture for the acquisition and reconstruction of hyperspectral images. The acquisition uses Compressed Sensing to separately acquire spectral rows, in the same way as actual satellite pushbroom sensors operate. The reconstruction relies on the minimization of Total Variation and a progressive refinement based on linear predictors to jointly process the measurements of each spectral row, in order to exploit both spectral and spatial correlation at the same time, with manageable complexity. Experiments run on AVIRIS and AIRS images show that the proposed approach, allowing to work with larger windows due to lower complexity with respect to existing algorithms, achieves similar performance as simpler but infeasible conventional approaches.



\clearpage
\bibliographystyle{IEEEbib}
\bibliography{jetcas_cs}

\end{document}